\documentclass[cits]{PoS}

\usepackage{amssymb,fontenc,times,mathptmx,graphicx}

\title{Exclusive processes in $pp$ collisions in CMS}

\ShortTitle{Exclusive processes in pp collisions in CMS}

\author{\speaker{Gustavo SILVEIRA, for the CMS Collaboration}%
         \thanks{This work is supported by CNPq/Brazil.}\\
        (Universite Catholique de Louvain (BE))\\
        E-mail: \email{gustavo.silveira@cern.ch}}

\abstract{
We report the results on the searches of exclusive production of low- and high-mass pairs with the Compact Muon Solenoid (CMS) detector in proton-proton 
collisions at $\sqrt{s}$~=~7~TeV. The analyses comprise the central exclusive $\gamma\gamma$ production, the exclusive two-photon production of dileptons, 
$e^{+}e^{-}$ and $\mu^{+}\mu^{-}$, and the exclusive two-photon production of $W$ pairs in the asymmetric $e^{\pm}\mu^{\mp}$ decay channel. No diphotons 
candidates are observed in data and an upper limit on the cross section is set to 1.18~pb with 95\% confidence level for $E_{T}(\gamma)>$~5.5~GeV and 
$|\eta(\gamma)|<$~2.5. In the measurement of exclusive dilepton production, seventeen $e^{+}e^{-}$ candidates are observed in data with $E_{T}(e)>$~5.5~GeV 
and $|\eta(e)|<$~2.5, and the cross section for the exclusive dimuon production is set to $\sigma(pp\to p\mu^{+}\mu^{-}p)=3.38^{+0.58}_{-0.55}$~(stat.)~$\pm$~0.16~(syst.)~$\pm$~0.14~(lumi.)~pb 
for $m(\mu^{+}\mu^{-})>$~11.5~GeV, $p_{\textrm{\footnotesize  T}}(\mu)>$~4~GeV and $|\eta(\mu)|<$~2.1, both in agreement with the theoretical predictions.
For high-mass systems, two events are observed in data for the exclusive two-photon production of $W$ pairs for $p_{\textrm{\footnotesize  T}}(\mu)>$~4~GeV, $|\eta(\mu)|<$~2.4 and 
$m(e^{\pm}\mu^{\mp})>$~20~GeV. Moreover, the study of the tail of the dilepton transverse momentum distribution resulted in model-independent upper limits 
for the anomalous quartic gauge couplings, which are of the order of 10$^{-4}$.
}

\FullConference{XXI International Workshop on Deep-Inelastic Scattering and Related Subjects\\
                 22-26 April, 2013\\
                 Marseilles, France}

\begin{document}

\section{Introduction}
\label{intro}

The mechanisms of exclusive production in proton-proton collisions, $pp\to p+X+p$, are characterized by the scattering of protons in small angles 
and the production of a central system with large rapidity gaps. In such interactions, no hadronic activity is observed in the final state apart of 
the particles in central rapidity and the very-forward protons. Considering these experimental signatures, the possible production mechanisms are  
the Pomeron ($I\!\!P$) and photon exchanges. The $I\!\!PI\!\!P$ interaction, which represents a color-singlet state usually described by the two-gluon 
exchange in the $t$-channel, is mostly applied for the diphoton and Higgs boson production~\cite{0111078}. Another possibility is the $\gamma I\!\!P$ 
interaction in view of the production of vector mesons~\cite{11063036,08052113}) and neutral vector bosons~\cite{08052113,9902481}. Also, the 
$\gamma\gamma$ interaction is employed for the production of pairs ($\gamma\gamma\to\ell^{+}\ell^{-},W^{+}W^{-}$) and is a well-known 
process in the framework of quantum electrodynamics (QED) with accuracy higher than 1\%. In this paper we report the results on the central 
exclusive production of diphotons~\cite{12091666}, and on the two-photon production of dileptons ($e^{+}e^{-}$~\cite{12091666} and $\mu^{+}\mu^{-}$~\cite{11115536}) 
and $W$ pairs~\cite{13055596} in proton-proton collisions at $\sqrt{s}$~=~7~TeV.%, where Fig.~\ref{fig:diagrams} presents the diagrams for each of these production processes. 

Considering that the forward protons escape the detector, large contributions come from processes with inelastic interactions with one of both 
protons dissociating into a low-mass hadronic system. Moreover, these processes are less understood theoretically and corrections due to rescattering have 
to be taken into account. Models are being developed to account for these corrections, often called survival probability, resulting in a survival probability close to 
100\%~\cite{0010163} for two-photon processes. On the other hand, these corrections are large for the central exclusive production of diphotons and are taken into 
account for the estimation of the cross section. Then, the contributions from elastic and inelastic interactions are included in the study of both exclusive production 
of dileptons and $W$ pairs.

\section{Diphoton production}
\label{diphoton}

The predictions obtained with {{\textsc{ExHuMe}} generator~\cite{exhume} are based in the {{\textsc{KMR}} model for the central exclusive production of diphotons~\cite{0111078}. In this model, 
the Pomeron is described by the two-gluon exchange in the $t$-channel, where two gluons participate in the hard interaction while a third soft gluon is exchanged between the protons 
to neutralize the color flow. Hence, the gluon emission from the proton is accounted with a parton distribution functions (PDF), which results in a sensitivity of the 
cross section to the low-$x$ gluon densities, namely $\sigma\sim [g(x)]^{4}$, where $x$ is the fraction of momentum of the proton carried by the gluon.

A data sample corresponding to a integrated luminosity of 36~pb$^{-1}$ is analyzed for the search of central exclusive diphoton events. The offline selection 
criteria requires exactly two photon candidates, each with $E_{T}(\gamma)>$~5.5~GeV and $\eta(\gamma)<$~2.5. Additionally, the photon candidates are 
required to pass the identification criteria with information from both HCAL and ECAL. As an electron can be misidentified as a photon, the hits in the pixel 
trackers are analyzed to account for this effect. In order to select only exclusive events, the offline selection also reject the tracks from the proton dissociation in the 
range $|\eta|<$~5.2. Due to the softness of the process, the two photon in the final state are expected to be back-to-back ($\Delta\phi\sim\pi$) and 
balanced in transverse energy ($\Delta E_{T}\sim 0$). Other sources of diphoton events are considered, such that non-exclusive production, exclusive $e^{+}e^{-}$ production (with an electron 
being misidentified as a photon), cosmic ray events and $\pi^{0}\pi^{0}$ production (with $\pi^{0}\to\gamma\gamma$). The expected number of background events is 1.79~$\pm$~0.40.

After applying all the selection criteria, no diphoton event is observed. The $CL_{s}$ approach~\cite{cls} is used to compute an upper 
limit for the production cross section, resulting in a limit with 95\% confidence level:
\begin{eqnarray}
\sigma[E_{T}(\gamma)>5.5\textrm{ GeV},|\eta(\gamma)|<2.5] < 1.18\textrm{ pb}.
\end{eqnarray}
This result includes both exclusive and semi-exclusive $\gamma\gamma$ production cross sections, with no particle activity from the proton dissociation within 
$|\eta|<$~5.2. Figure~\ref{fig:diphoton-epem} (left) presents the estimated upper limit in comparison to different theoretical predictions for the central exclusive diphoton 
production (only the elastic process). %The predictions comprise both MRST and MSTW PDFs with leading order (LO) and next-to-leading order (NLO) in the strong coupling 
%constant, $\alpha_{s}$.
As a result, the estimated upper limit is one order of magnitude above the predictions with next-to-leading order (NLO) PDFs.

\begin{figure}
\includegraphics[width=.45\textwidth]{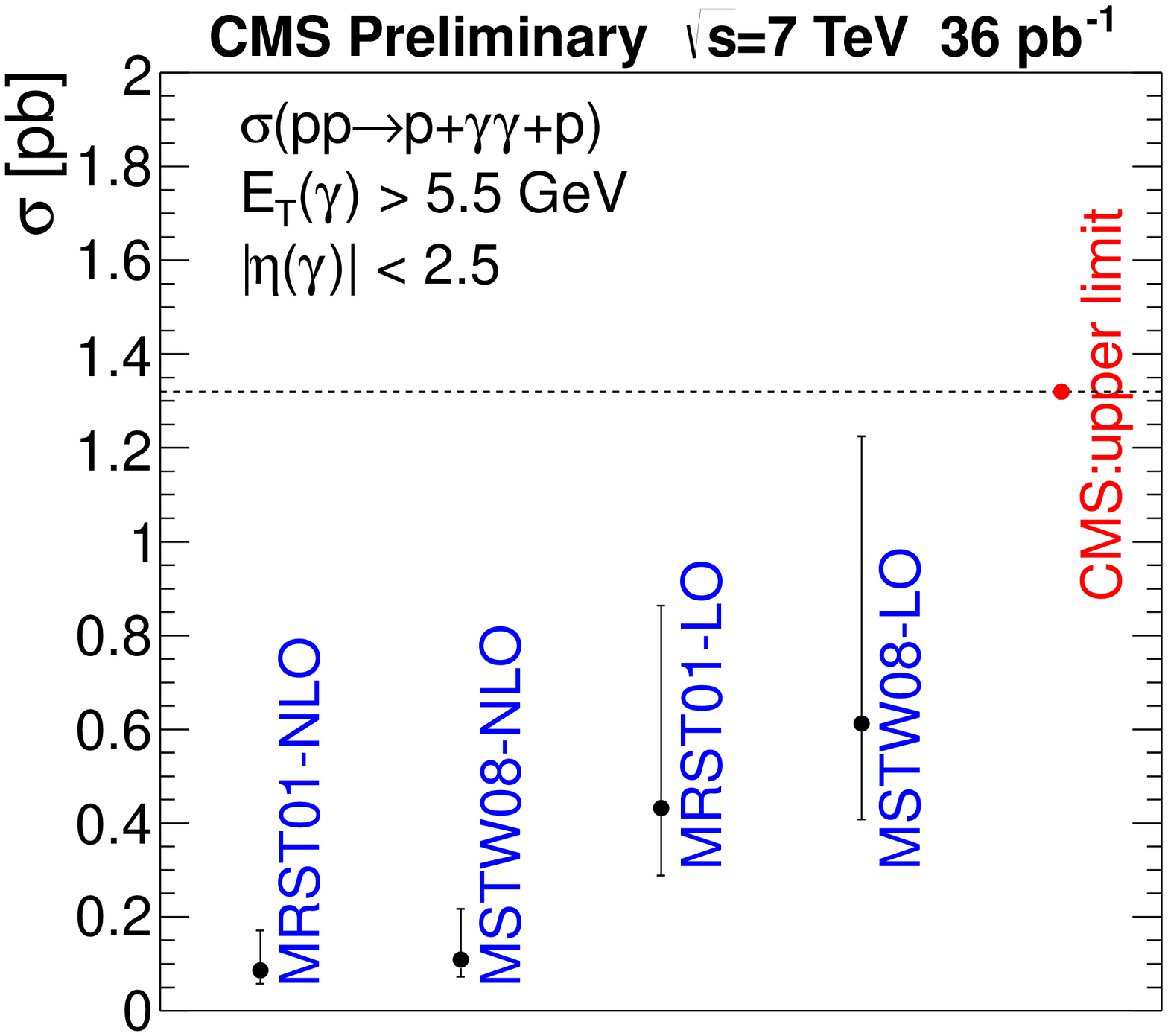}
\hspace{1em}
\includegraphics[width=.45\textwidth]{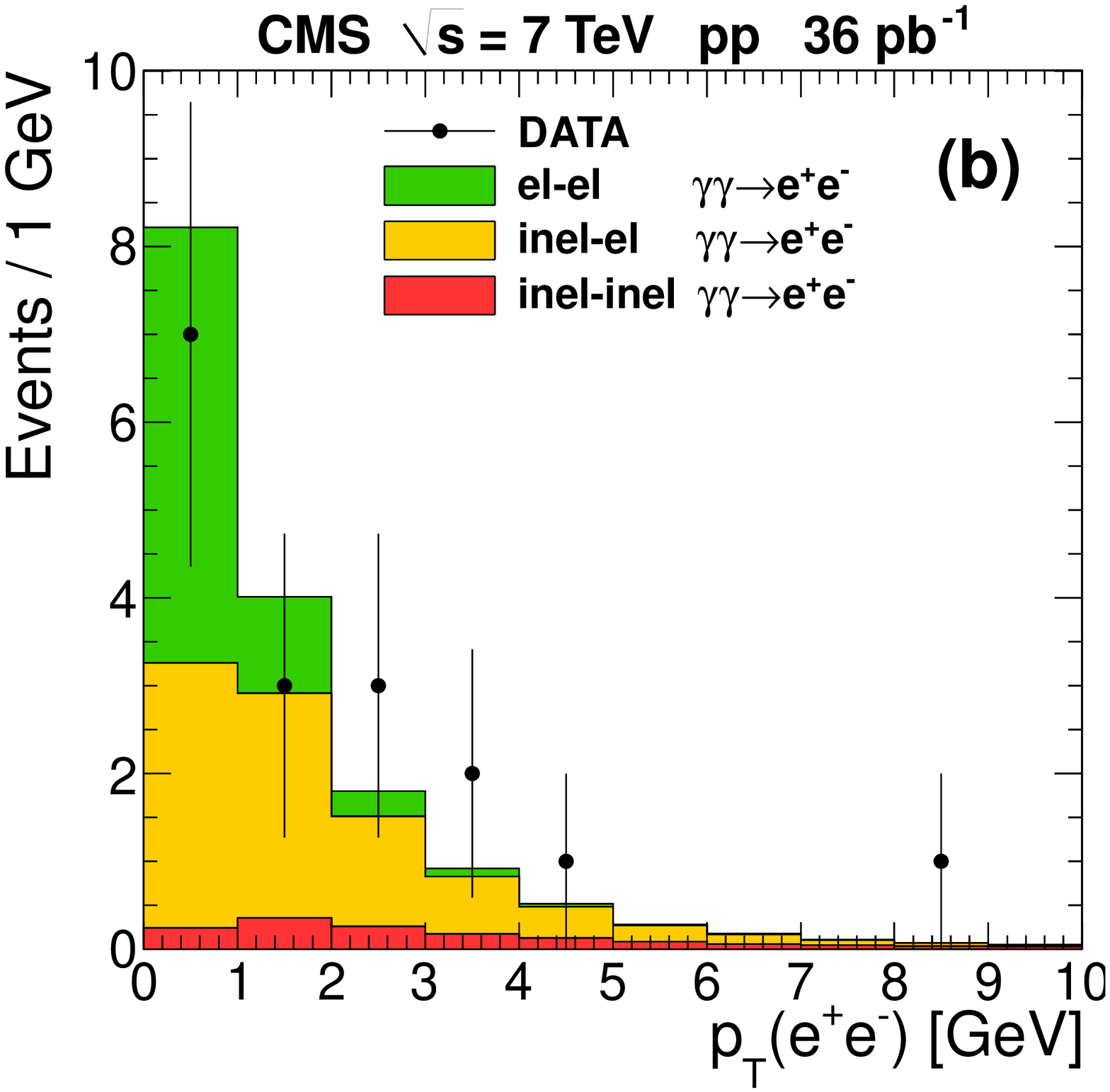}
\caption{\label{fig:diphoton-epem}
(Left) Comparison of the estimated upper limit for the central exclusive diphoton production with theoretical predictions. (Right) Transverse momentum distribution of the 
$e^{+}e^{-}$ pair.
}
\end{figure}

\section{Dilepton production}
\label{dileptons}

Following the same procedure as in the diphoton case, the analysis of the exclusive production of dileptons only events with no additional tracks associated to the dilepton 
vertex. In the case of $e^{+}e^{-}$ production, a data sample containing an integrated luminosity of 36~pb$^{-1}$ is used to select exactly two electron tracks 
with opposite charges and with $E_{T}(e)>$~5.5~GeV and $\eta(e)<$~2.5. The analysis also take care 
of background events associated to the non-exclusive $e^{+}e^{-}$ production (controlled by the number of additional 
tracks associated to the vertex), exclusive $\Upsilon$(nS) production ($\gamma I\!\!P\to\Upsilon$(nS)$\to e^{+}e^{-}$%~\cite{0311164}
and completely rejected with $E_{T}(e)>$~5.5~GeV), cosmic ray contamination, and exclusive $\pi^{+}\pi^{-}$ production via $I\!\!PI\!\!P$ fusion. It follows that the expected 
number of background events is 0.85~$\pm$~0.28.

After electron reconstruction and identification, cosmic-ray rejection and exclusivity requirements, seventeen events are observed in data in 
comparison to 16.3~$\pm$~1.3~(syst.) events expected from theory. Figure~\ref{fig:diphoton-epem} (right) presents the $p_{\textrm{\tiny T}}$ distribution of the $e^{+}e^{-}$ candidates in 
agreement with electrons being back-to-back ($\Delta\phi\sim\pi$) and balanced in transverse energy ($\Delta E_{T}\sim 0$).

The two-photon production of $\mu^{+}\mu^{-}$ is studied in a event sample with an integrated luminosity of 40~pb$^{-1}$. Similarly, events with exactly 
two muon tracks are selected in data containing no other tracks associated to the dimuon vertex. 
In addition, vertices related to exclusive dimuon events are separated from those related to the background, like by Drell-Yan (DY) and QCD production, with the requirement 
of no additional tracks be separated by less than 2~mm from the dimuon vertex.

The exclusivity selection for the dimuons follows the requirements to keep higher muon efficiencies, then only muons with $p_{\textrm{\footnotesize  T}}(\mu)>$~4~GeV and 
$|\eta(\mu)|<$~2.1 are retained. Also, the requirement of $m(\mu^{+}\mu^{-})>$~11.5~GeV rejects events from the exclusive $\Upsilon$(nS) production, 
$\gamma I\!\!P\to\Upsilon$(nS)$\to \mu^{+}\mu^{-}$, 
and selecting events with opening angle smaller than 0.95$\pi$ reduces the contribution from cosmic ray muons. Finally, muon candidates are required to be back-to-back, 
related to the acoplanarity cut of $1-|\Delta\phi(\mu^{+}\mu^{-})/\pi|<$~0.1, and balanced in $p_{\textrm{\tiny T}}$, 
$|\Delta p_{\textrm{\footnotesize  T}}(\mu^{+}\mu^{-})| = \vec{p}_{T}(\mu^{+}) - \vec{p}_{T}(\mu^{-})|<$~1.0~GeV. 
The final result is 148 events passing all the criteria. Figure~\ref{fig:dimuons} shows 
the $p_{\textrm{\tiny T}}$ and $\Delta p_{\textrm{\footnotesize  T}}$ distributions of the dimuons considering elastic and inelastic production with background events from the 
DY production. As a result, the measured cross section is set to:
\begin{eqnarray}
\sigma(pp\to p\mu^{+}\mu^{-} p) = 3.38^{+0.58}_{-0.55}\textrm{ (stat.) }\pm \, 0.16\textrm{ (syst.) }\pm \, 0.14\textrm{ (lumi.) pb,}
\end{eqnarray}
which is consistent with the theoretical QED predictions~\cite{lpair}.

\begin{figure}
\includegraphics[width=.45\textwidth]{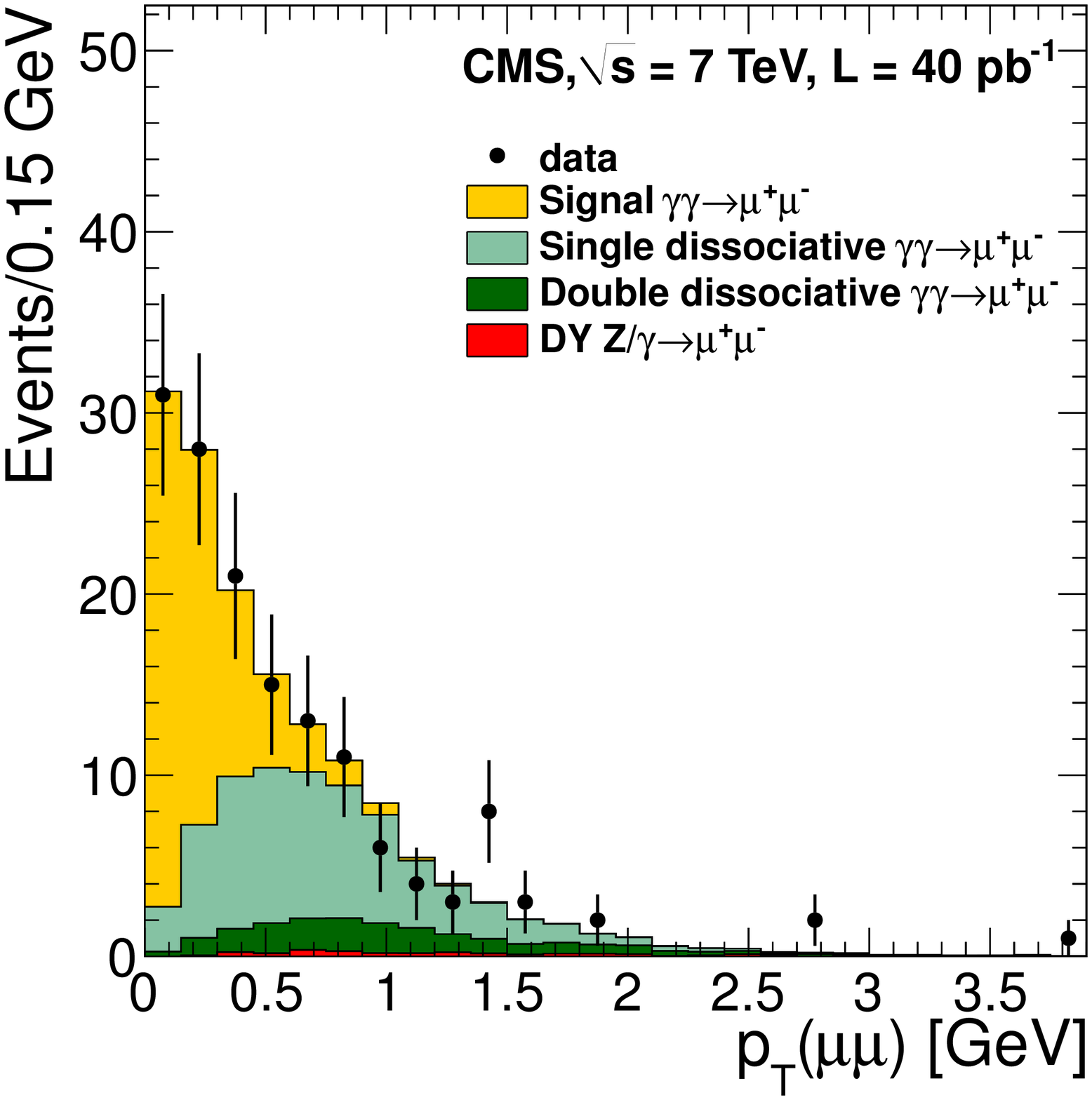}
\hspace{1em}
\includegraphics[width=.45\textwidth]{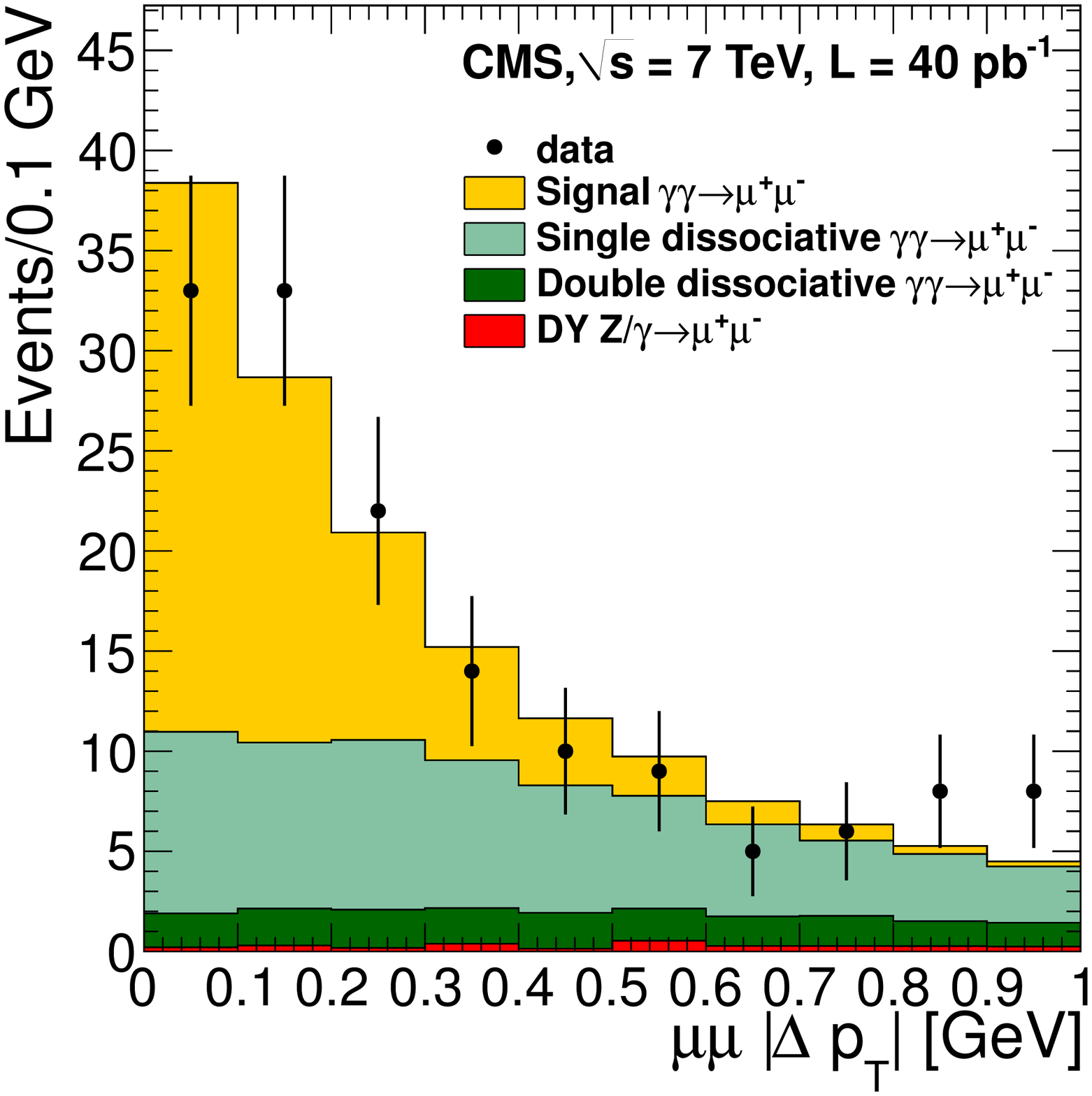}
\caption{\label{fig:dimuons}
Transverse momentum and $\Delta p_{\textrm{\footnotesize  T}}$ distributions for the dimuons.
}
\end{figure}

\section{$W$ pairs}
\label{wpairs}

The search for exclusive two-photon production of $W$ pairs analyzes the processes related to the standard model (SM) diagrams with the single $\gamma\gamma\to W^{+}W^{-}$ 
coupling and those with $W$ exchange in the $t$- and $u$-channels~\cite{13055596}. A data sample corresponding to 5.05~fb$^{-1}$ is analyzed to select exclusive events where the 
$W$ bosons decay into charged leptons, following a similar procedure as in the dimuon case. In order to reduce the contribution of background events from the DY 
production of dimuons, the search for exclusive $W$-pair events is carried out in the $e^{\pm}\mu^{\mp}$ decay channel with asymmetric triggers with $p_{\textrm{\footnotesize T}}(\ell)$ thresholds 
of 17~GeV and 8~GeV. The selection criteria consists of leptons with $p_{\textrm{\footnotesize  T}}(\ell)>$~20~GeV and $|\eta(\ell)|<$~2.4 and pairs with an invariant mass larger than 20~GeV. Then, 
the exclusivity requirement ensures that events have no additional tracks related to the dilepton vertex, acoplanarity of $1-|\Delta\phi(\ell^{+}\ell^{-})/\pi|<$~0.1 and 
$|\Delta p_{\textrm{\footnotesize  T}}(\ell^{+}\ell^{-})|<$~1.0~GeV. Moreover, a kinematic cut of $p_{\textrm{\footnotesize  T}}(\ell^{\pm}\ell^{\mp})>$~30~GeV is used to reduce the contribution from DY production of 
dimuons and $\tau^{+}\tau^{-}$.

Besides, the $\gamma\gamma\to\mu^{+}\mu^{-}$ process is still studied in order to keep a control sample to validate the selection and estimate the 
proton dissociative contribution, since no event generator is available for the semi-exclusive production $W$ pairs. The {{\textsc{Lpair}} generator~\cite{lpair} is
employed to produce the simulated event samples for the elastic and inelastic production of dimuons. In this study, a scale factor for the proton dissociative contribution 
is estimated as $F$~=~3.23~$\pm$~0.50~(stat.)~$\pm$~0.36~(syst.) in the region of $m(\mu^{+}\mu^{-})>$~160~GeV~\cite{13055596} and used to scale the signal predictions for the $W^{+}W^{-}$ events.

After passing all the criteria, 2 events are observed in data in comparison to 2.2~$\pm$~0.5 signal and 0.84~$\pm$~0.13 background events expected. This result 
can be converted into a cross section for the exclusive two-photon production of $W$ pairs like:
\begin{eqnarray}
\sigma(pp\to p^{(*)}W^{+}W^{-} p^{(*)}\to p^{(*)}e^{\pm}\mu^{\mp} p^{(*)}) = 2.1^{+3.1}_{-1.9}\textrm{ fb,}
\end{eqnarray}
where $p^{(*)}$ denotes that the proton dissociative contribution is taken into account. Figure~\ref{fig:ww} (left) shows the $p_{\textrm{\tiny T}}$ distribution of 
the dimuons with the two observed events.

Furthermore, this analysis explore the possible contribution of anomalous couplings in the $\gamma\gamma\to W^{+}W^{-}$ vertex, since any deviation from the SM prediction may indicate 
an evidence of new physics. In this scenario, a modified model for the SM Lagrangian is used as input in the {{\textsc{CalcHEP}} generator~\cite{calchep}
where the parameters for the aQGC, $a^{W}_{0}$ and $a^{W}_{C}$, are included by dimension-6 operators~\cite{belanger}. However, this modification brings the violation of 
unitary at high energies, since the cross section rises quadratically with the center-of-mass energy of the $\gamma\gamma$ system, $W_{\gamma\gamma}$. In order to tame this 
rising, the anomalous parameters are weighted with a form factor of the form:
\begin{eqnarray}
a^{W}_{0,C} (W^{2}_{\gamma\gamma}) = \frac{a^{W}_{0,C}}{\left( 1 + \frac{W^{2}_{\gamma\gamma}}{\Lambda^{2}_{\textrm{\tiny cutoff}}} \right)^{p}},
\end{eqnarray}
where $p$ is set to 2.0 (dipole form factor) based in previous aQGC studies~\cite{d0anom} and $\Lambda_{\textrm{\scriptsize cutoff}}$ is the 
energy scale for new physics, which is set to 500~GeV. %Then, MC event samples are produced for a variety of anomalous parameters in order to compare with data.

The selection requirements are the same as applied for the SM case instead of a larger restriction for the $p_{\textrm{\tiny T}}$ of the lepton pair: a cut of 
$p_{\textrm{\footnotesize  T}}(\ell^{\pm}\ell^{\mp})>$~100~GeV is applied in order to suppress the SM contribution, which is accounted as background events. After all the selection 
criteria applied, no event is observed in data. In this case, an upper limit for the production cross section is estimated with 95\% confidence level:
\begin{eqnarray}
\sigma(pp\to p^{(*)}W^{+}W^{-} p^{(*)}\to p^{(*)}e^{\pm}\mu^{\mp} p^{(*)}) < 1.9\textrm{ fb}.
\end{eqnarray}
Additionally, limits on the anomalous parameters are estimated using the $CL_{s}$ method~\cite{cls}, resulting in:
\begin{eqnarray}
|a^{W}_{0}| < 2.8 \times 10^{-6}\textrm{ GeV}^{-2}, \, |a^{W}_{C}| < 1.0 \times 10^{-5}\textrm{ GeV}^{-2}
\end{eqnarray}
considering no form factor in the anomalous parameters, and
\begin{eqnarray}
|a^{W}_{0}| < 1.7 \times 10^{-4}\textrm{ GeV}^{-2}, \, |a^{W}_{C}| < 6.0 \times 10^{-4}\textrm{ GeV}^{-2}
\end{eqnarray}
including the form factors with the energy scale $\Lambda_{\textrm{\scriptsize cutoff}}=$~500~GeV. Figure~\ref{fig:ww} (right) presents the 2-dimensional limits in comparison to the 
expectation from the SM. As a result, these limits are two order of magnitude more stringent than those obtained in LEP~\cite{opal} due to the lower center-of-mass energy 
probed in the $e^{+}e^{-}$ collisions.

\begin{figure}
\includegraphics[width=.45\textwidth,angle=-90]{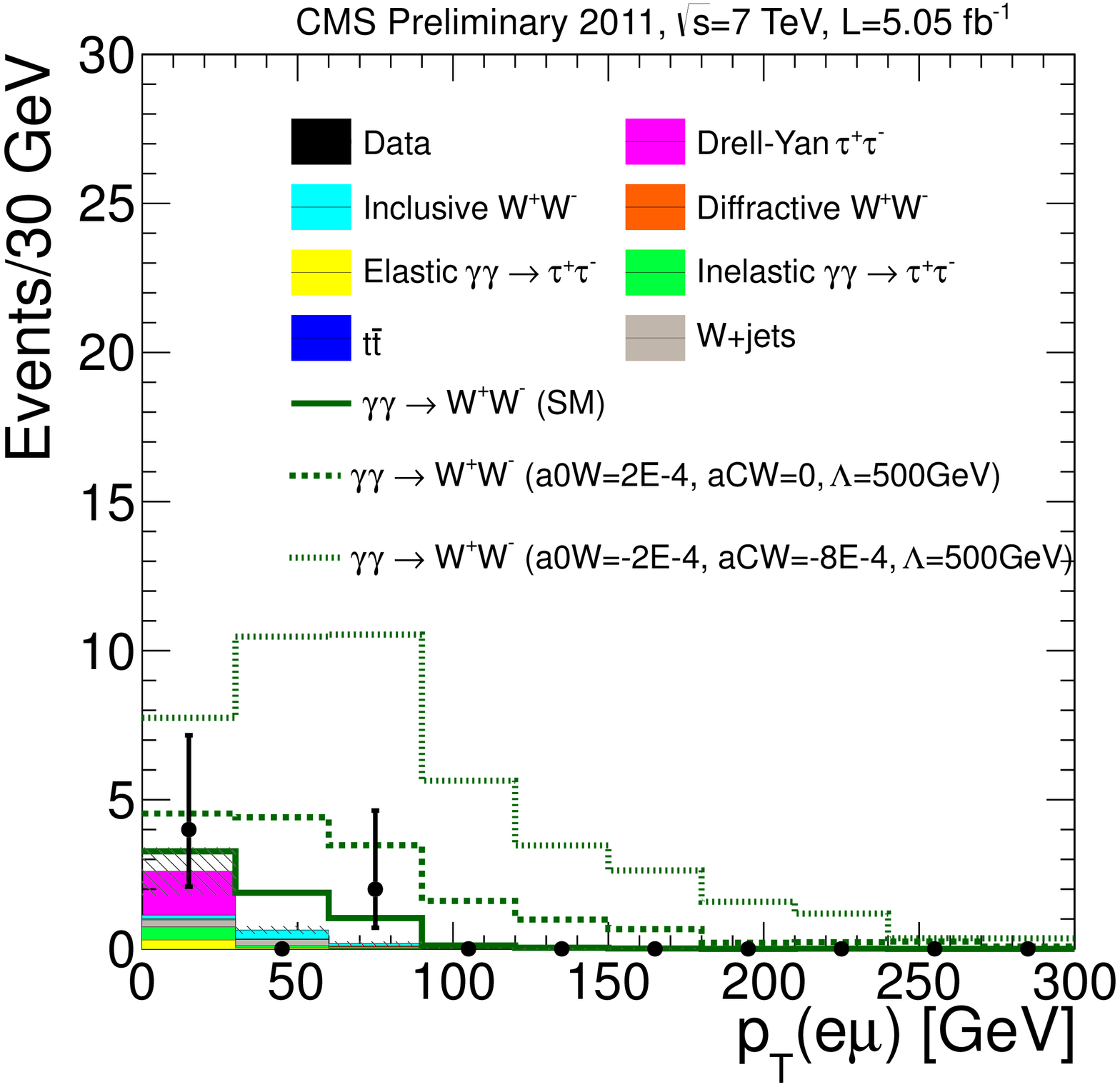}
\hspace{1em}
\includegraphics[width=.45\textwidth,angle=-90]{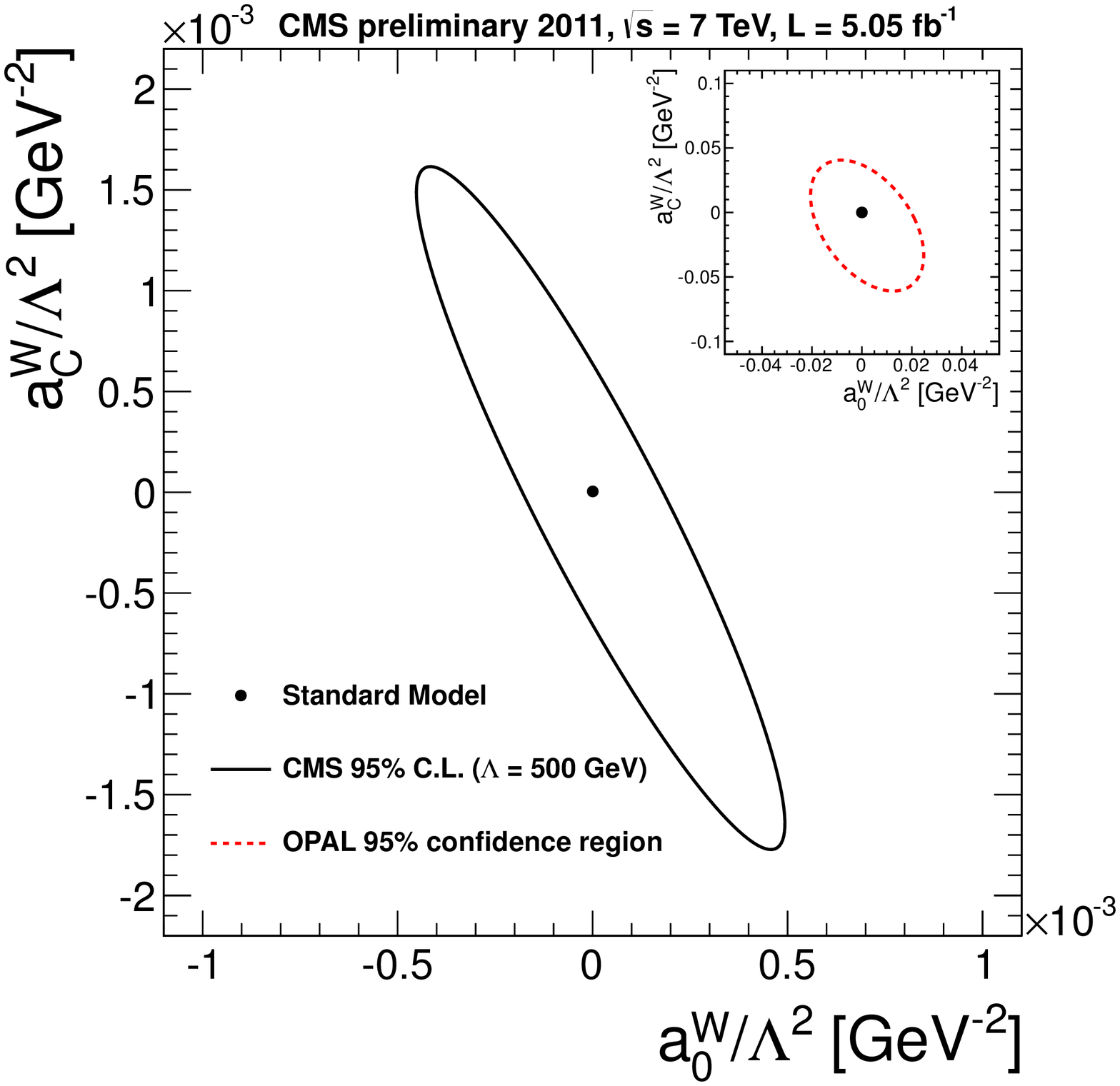}
\caption{\label{fig:ww}
(Left) Transverse momentum distribution of the $e^{\pm}\mu^{\mp}$ pair with the two observed candidates. (Right) Two-dimensional limit for the $a^{W}_{0}$ and $a^{W}_{C}$ 
anomalous parameters for $\Lambda_{\textrm{\scriptsize cutoff}}=$~500~GeV in comparison to the previous limits provided with the LEP data.
}
\end{figure}

\end{document}